# Host-Guest Interactions in $ExBox^{4+}$

***Ranjita Das and Pratim Kumar Chattaraj****

Department of Chemistry and Center for Theoretical Studies, Indian Institute of Technology

Kharagpur, Kharagpur – 721302, West Bengal, India

*Author for correspondence: Email: pkc@chem.iitkgp.ernet.in

**Abstract** The host-guest interaction between benzene/ azine with the newly synthesized $ExBox^{4+}$ complex is studied with the help of DFT. The solvent phase interaction energy is found to decrease with gradual substitution of methine group (=CH–) of guest benzene ring with N atom in the resultant azine@$ExBox^{4+}$ complex. The nature of bonding interaction is studied with the help of newly developed NCI plot program package along with energy decomposition analysis (EDA) and charge decomposition analysis (CDA). The interaction is mostly π-type van der Waals interaction.

## 1. Introduction

Supramolecular chemistry has immense contribution in the fields of nano-technology, green chemistry, designing catalyst, medicinal chemistry etc.[1,2] The invention of Crown ether by Perdersen[3] started a new age in supramolecular chemistry including "host-guest interaction". The newly synthesised $ExBox^{4+}$ by the Stoddart group[4-10] is the most recent addition to this field. $ExBox^{4+}$ comprises 1,4-phenylene-bridged bipyridinium units, and adopts a box like structure. $ExBox^{4+}$ acts as a scavenger for an array of polycyclic aromatic hydrocarbons (PAHs) in aqueous or organic medium. Recently a new variety, $Ex^2Box^{4+}$, is also synthesized that can bind π-electron rich as well as π-electron deficient hydrocarbons.[10] The $Ex^2Box^{4+}$ can also form complex with polyether macrocycles, like 1,5-dinaphtho[38]- crown-10.

In all these complexes the host interacts with the guest molecule through non-covalent interaction. A recent theoretical study reported that the energy of interaction between the guest and the host varies linearly with the size of the guest molecule up to anthracene.[11] The non-covalent interaction between guest and the host has been an interesting and challenging subject of research. Here the interaction between the $ExBox^{4+}$ and the guest PAH is expected to include π-π type interaction. In the present study we try to understand the nature of bonding interaction between the $ExBox^{4+}$ complex and the guest aromatic systems. In addition, the effect of solvent and counter-ion on the strength of bonding interaction is also studied. We consider benzene as the guest molecule.

We also try to understand the effect of inclusion of hetero atom in the aromatic hydrocarbon guest, on the bonding interaction. Some azines are considered as guest molecules. The studied heterocyclic aromatic compounds are isoelectronic with benzene. We try to figure out the variation of interaction energy with successive substitution of methine group (=CH–) from the guest benzene ring with N atom up to N=4. These heterocyclic molecules are of huge chemical and biological interest.[12-14] The azines are used as precursor to pharmaceutical, commercial resins etc. The derivatives of the azines are also used as important antibiotic and antitumor agents. Some of the azines are abundant in nature. Azines are also synthesized through some existing specialized methods. Owing to the tremendous use of these heterocyclic aromatic molecules, extraction of these azines is quite important. Here we present the nature of interaction of $ExBox^{4+}$ with these heterocyclic aromatic molecules and compare them with isoelectronic aromatic hydrocarbons.

## Computational details

The geometries of the guest-host complexes are optimized at wB97x-D /6-311G(d, p) level. The frequency calculation is also performed at the same level of theory. In order to get an accurate geometry and good interaction energy the free optimization of the host-guest complexes and frequency calculation are performed using wB97x-D method.[15] The wB97x-D is the latest functional from Head-Gordon and co-workers, which includes empirical dispersion as well as long range corrections. In order to understand the effect of solvent on the interaction energy, optimization is also carried out in solvent phase adopting a Conductor-like Polarizable Continuum Model (CPCM)[16] at the same level of theory using water and acetonitrile as solvents. Nucleus independent chemical shift (NICS)[17-19] values of the guest molecules are calculated at the same level of theory. All these calculations are performed using G09 program package.[20]

Some selected structures are re-optimized at B3LYP-D/6-311G (d, p) level. Energy decomposition analysis (EDA) is performed with the help of ADF program package[21] by importing the B3LYP-D optimized geometry. EDA calculation is performed by expanding the MOs in an uncontracted set of STO containing double zeta polarization functions (DZP). The nature of charge decomposition is analysed by adopting charge decomposition analysis (CDA)[22] scheme as implemented in Multiwfn program.[23]

Recently introduced NCI program package[24] by Yang and co-workers is used to plot and explore the non-covalent interactions (NCI)[24-26] only from the geometrical knowledge of molecule. NCI uses two scalar quantities electron density, $\rho$, and reduced density gradient (RDG, s) to map the localized bonding interaction. The reduced density gradient (RDG, s) is defined as

$$\boldsymbol{s} = \frac{1}{2(3\pi)^{1/3}}\frac{|\nabla\rho|}{\rho^{4/3}} \ldots\ldots\ldots\ldots\ldots (1)$$

where $\nabla\rho$ signifies the gradient of the electron density, $\rho$. The RDG assumes a large value in the region away from the molecule, while it adopts a very small value, nearly 0, in the regions of covalent and/or non-covalent bonding. The color scheme is a red-green-blue scale where the attractive interaction ($\rho_{cut}^{-}$) is described by blue color, while red signifies repulsive interaction ( $\rho_{cut}^{+}$). Recently Yang group provided an algorithm to compute NCI in order to visualize and evaluate the weak non covalent interactions.[24]

## 2. Results and discussion

### 2.1. Structure and geometry

Here we consider that the host and guest molecules interact through π-π type interaction. We already know that there are three possible conformations for π-π stacking, and only two conformations are energetically stable, the T-shaped conformation and parallel displaced conformation. Hence it is expected that during interacting with the six-membered rings of $ExBox^{4+}$ the guest molecule will prefer to adopt either of these two conformations. Previous reports suggest that in the resultant inclusion compound, due to small size, these six-membered aromatic guest compounds will prefer to occupy either side of the $ExBox^{4+}$. The previously reported lowest energy structure for Benzene@$ExBox^{4+}$ (in gas as well as solution phase) is considered.[11] We optimize different geometries of Guest@ $ExBox^{4+}$ complex for each of the guest and only the geometries with NIMAG equals to zero are presented in Figure S1. For Pyridine@$ExBox^{4+}$ complex we found three different geometries. The structure with the lowest energy among them is shown in Figure 1 and the rest are displayed in Figure S1. All these co-conformers belong to minima on the PES and are close in energy. The pyridine ring adopts a parallel displaced conformation with pyridinium rings (top and bottom chains) and T-shaped conformation with the phenyl ring (side chain). The other two conformations are 0.5kcal/mol and 0.4kcal/mol above the lowest energy structure. In order to observe the effect of solvent, the host-guest structures are re-optimized using CPCM model for two solvents, i.e. acetonitrile and water. The solution phase structure is in good agreement with gas phase geometry (Figure S2).

For Pyrimidine@$ExBox^{4+}$ complex we found total four structures. Figure 1 shows that the guest pyrimidine ring encounters with pyridinium rings (top and bottom chains) in T-shaped manner and with the phenyl ring (side chain) it interacts in parallel displaced manner. The structures with pyrimidine ring placed in T-shaped manner to the phenyl ring are 3.2-1.5 kcal/mol less stable than the previous structure. In solution phase the structure is quite different, the 2nd lowest energy structure becomes the lowest energy structure in solution phase. Starting the solution phase optimization with the geometry of lowest energy structure as initial guess also leads to the same geometry of the 2nd lowest energy structure (Figure S2).

For Pyridazine@$ExBox^{4+}$ complex, five possible structures are found. Among them the lowest energy structure attains the same conformation as that of pyridine guest molecule. The solution phase structures also follow same relative energy order as that of the gas phase calculation. Three geometries are found for Pyrazine@$ExBox^{4+}$ complex in which the guest molecule is positioned in a parallel displaced manner to the pyridinium rings. The lowest energy structure among them is provided in Figure 1. In solution phase the second lowest energy structure in gas phase becomes the lowest energy structure. For the guest molecules with three N atoms in the ring we found only two structures of guest-host complex for each guest at the wB97-XD/6-311G (d, p) level of theory. But we found some other structures at lower level of theory (wB97-XD/6-31G (d, p)) but upon optimization on higher basis set either the structures acquire the structure shown in Figure 1 or a structure with imaginary frequency. It is observed that in 1,3,5Triazine@$ExBox^{4+}$ complex the guest ring favors to be placed in a parallel displaced manner to the pyridinium rings (top and bottom chains) and T-shaped fashion to the phenyl ring (side chain). The host cage gets severely damaged if we place the 1,3,5 Triazine ring parallel to phenyl ring. The solution phase structure also adopts the same conformation. In case of 1,2,3, Triazine@$ExBox^{4+}$ and 1,2,4 Triazine@$ExBox^{4+}$ complexes the guest is placed in parallel displaced manner with the phenyl ring (side chain). The solvent phase structures are in good agreement with those in gas phase for 1,2,4 triazine, whereas for 1,2,3 triazine the second co-conformer is energetically more stable.

For the guest molecules having four N atoms (tetrazines), it is observed that the guest-host complexes are more stable when the guest molecule is positioned in parallel displaced conformation to the phenyl ring (side chain). For 1,2,3,4Tetrazine@$ExBox^{4+}$ complex we found another structure close in energy to the one with lowest energy structure, the guest molecule is placed in parallel displaced conformation to the phenyl ring. In solution phase the lowest energy structures are identical with gas phase lowest energy structure. For 1,2,4,5Tetrazine@$ExBox^{4+}$ we found only one structure with NIMAG equals to zero in gas phase calculation. The solution phase structure is slightly different than that of gas phase structure. The geometrical parameter for Guest@ $ExBox^{4+}$ complexes are provided in table S1.

It is observed that in solution phase generally the guest molecules (no of N atom=0-3) prefers to adopt a parallel displaced conformation to interact with the pyridinium rings except for 1,2,4 triazine. Tetrazine prefers to adopt a T-shaped conformation to interact with the phenyl ring.

In gas phase the HOMO-LUMO gap (HLG) of Benzene@$ExBox^{4+}$ complex is 7.573 *eV* which suggests the stability of the complex. Upon replacing the benzene with N substituted azabenzenes the HLG of the Guest@ $ExBox^{4+}$ complexes becomes quite close even in some cases higher than that of the Benzene@$ExBox^{4+}$ complexes implying stability of other Guest@ $ExBox^{4+}$ complexes.

## 2.2. Interaction energy

In gas phase calculation the computed interaction energy for Benzene@$ExBox^{4+}$ complex appears to be -19.9 kcal/mol. The reaction enthalpy suggests that the process is exothermic in nature, and the negative reaction free energy implies the thermodynamic spontaneity of the process and the result corroborates well with the previous investigation.[11] The interaction energy for other Guest@ $ExBox^{4+}$ complexes, while the guests are N substituted benzenes, is listed in Table 1. In gas phase calculation the interaction energy varies within the range of 21.6-17.0 kcal//mol.

Since the interaction is considered as of π-type, the π-electron density of the guest molecule should play a key role. Although these heterocyclic compounds are isoelectronic with the benzene ring the even distribution of electron density over the ring is disrupted by the presence of N atom compared to the benzene ring and that causes a decrease in resonance energy in heterocyclic compounds compared to that in benzene ring. Hence stability as well as aromaticity of these guest molecules are important factors that may influence the extent of interaction. The stability of these guest compounds is closely related to their aromaticity. But the aromaticity of azines has been a subject of debate for past few years. Mandado et al. revealed[27] that for azines the stability order is not only controlled by aromaticity criteria but also structural factors. The aromaticity order as per NICS(0) value corroborates well with the stability order for diazines, But for tetrazine and triazine the least stable is the most aromatic according to NICS(0) value. They suggested that the aromaticity should decrease on insertion of N-atom unless a direct N-N bond exists. Later Solà and co-worker reported[28] that with increasing substitution of (=CH–) group with N atom the aromaticity of hetero cyclic planar compound decreases. They also reported that for a particular number of N atoms the ring with largest number of N-N bonds is most aromatic. They found that NICS values do not provide any clear idea of their aromaticity order. On contrary to these reports recently Wang et al suggested[29] that the insertion of N atom should not decrease the aromaticity of six membered guest molecules compared to that of benzene ring. Like these reports we also found that the calculated NICS value is sensitive towards the level of theory used in the calculation. In this present study we try to analyse how insertion of hetero atom, i.e. N atom, in benzene ring influences the host-guest interaction. We have calculated NICS(0) for the guest molecule and found that on successive insertion of N atom the NICS(0) value decreases but we do not see any trend for $NICS(0)_{zz}$ value.  We try to understand the influence of aromaticity of the guest molecule on the intermolecular host-guest interaction. In gas phase no regular trend is observed in interaction energy on successive insertion of N atom to the guest molecule.

The solvent phase interaction energy and interaction energy in presence of counter-ion (in gas phase) listed in tables 2 and 3. In presence of counter ion, i.e. chloride ions, we observe that there occurs a gradual decrease in interaction energy of the Guest@$ExBox^{4+}.4Cl^{-}$ complexes with increasing number of N atoms in the guest ring (for N = 0-4) (Figure 2). Again in presence of solvent the interaction energy of the Guest@$ExBox^{4+}$ is found to decrease regularly (with some exception) with increasing number of N atoms in heterocyclic aromatic ring (Figure 2). We also observed that the interaction energy in solution phase gradually decreases with decreasing the NICS(0) value (absolute value) of the guest molecule (Figure S3). Even in presence of counter-ion the interaction energy gradually decreases with decreasing NICS(0) value with few exceptions. In solvent phase the interaction energy for N=3 systems shows a little variation by changing the guest molecule ( or the NICS value of guest molecule).

It is observed that in presence of counter-ion the interaction between the guest molecule and host ($ExBox^{4+}.4Cl^{-}$) becomes stronger and the reaction free energy values are higher than that of the Guest@$ExBox^{4+}$complexes (table 3). But in presence of solvent the interaction between the guest molecule and host ($ExBox^{4+}$) becomes weaker compared to that of gas phase calculation. It is observed that the reaction free energy (ΔG) gradually decreases with increasing number of the N atom in the heterocyclic ring, even for 1,2,4 triazine and tetrazine molecules as guest the ΔG becomes positive (table 2) suggesting that the complexation process is thermodynamically unfavorable at room temperature and for spontaneous reaction somewhat lower temperature is required. Figure 2 suggests that the trend of interaction is same in presence of both acetonitrile and water as solvent. The effect of solvationon the interaction energy and ΔG is in agreement with the previous theoretical study. The energy gained by the interaction between the host molecule and the guest molecule is partly compromised by the loss of solvation energy. It is possible to get a rough idea about the type of interaction from the interaction energy value and here we see that the interaction is non-covalent type. Further the decrease in interaction energy in solution phase also suggests that the interaction is of non-covalent type. In the following section the nature of interaction is studied thoroughly.

## 2.3. Nature of interaction:

Here we analyse the bonding interaction between the guest molecule and host complex. The analysis is performed on the gas phase structures since the attractive interaction is most distinct in gas phase rather than in solution phase. The Guest@$ExBox^{4+}$ complexes are analysed for bonding interaction. The type of interaction is analysed by means of bond energy decomposition analysis (BEDA) and non-covalent interaction (NCI) plot.

The NCI analysis of Guest@ExBox$^{4+}$ complexes reveals the nature of intermolecular interaction between the guest and the host molecules. The NCI plot allows us to identify the interacting regions as well as assessment of the type of interaction. Figure 3 shows the guest molecules interacting with the pyridinium ring of top and bottom chains and phenyl ring of the side chain of the host complex either in T-shaped or in parallel displaced manner. The broad intermolecular isosurface of NCI plot suggests that an attractive van der Waals type interaction (Figure S4) is operative between the guest and host molecules. The values of ρ near zero indicates the van der Waals interaction between the guest and the host molecules.

The literature suggests that π-interaction comprises balanced contribution of quadrupole-quadrupole interaction and London dispersion interaction, even some times charge transfer interaction is a component.[30-32] To analyse the interaction type in detail we performed BEDA and also CDA. The BEDA data are provided in table 4. It is found that for Benzene@ ExBox$^{4+}$ complex the electrostatic energy ($\Delta V_{eles}$), orbital interaction energy ($\Delta E_{orb}$) and dispersion interaction energy ($\Delta E_{dis}$) have significant contributions to the total attractive interaction term. It is observed that there is a good competition between the $\Delta V_{eles}$ and $\Delta E_{dis}$ terms. The EDA suggests that the interaction is of non-covalent type, mostly of π-type interaction. For all azine@ExBox$^{4+}$ complexes the contributions from electrostatic and dispersion interactions dominate over the contribution from orbital interaction. It is observed that the contribution of different energy terms is influenced by the mode of interaction between the azine and host complex. It is observed that the contribution of electrostatic interaction energy dominates over other components when the guest molecule interacts in a T-shaped manner with the pyridinium ring of ExBox$^{4+}$. As in the case of pyrimidine, 1,2,3, triazine, 1,2,4 triazine, 1,2,3,4 tetrazine, 1,2,4,5 tetrazine guest molecules the contribution of electrostatic interaction energy is more or less 50%. In T-shaped conformation the quadrupole-quadrupole interaction is favorable, and eventually decreases the contribtion of dispersion terms.

The CDA indicates the direction as well as extent of electron flow within the Guest-Host complex (table 5). It is observed that there occurs an electron flow from the guest molecule to the host. A careful scrutiny of the CDA data along with the structure of the complex suggests that the extent of charge transfer is dominant for the structures where the guest molecule interacts with the phenyl ring of side chain in a parallel displaced manner and adopts a T-shaped conformation with the pyridinium ring of ExBox. However for all these structures the charge transfer interaction is not very prominent, rather van der Waals interaction surpasses the charge transfer interaction.

### 2.4. 2Guest@ExBox$^{4+}$.4Cl$^{-}$ complex

Next we add a second molecule of guest to the Guest@ExBox$^{4+}$.4Cl$^{-}$ complex in order to find out whether the host, i.e. ExBox$^{4+}$.4Cl$^{-}$ complexes, can hold two molecules of guest or not. The calculated HLG, interaction energy and thermodynamic parameters (both sequential and average) are listed in table 5. The HLG is substantially high to indicate the stability of the resultant 2Guest@ExBox$^{4+}$.4Cl$^{-}$ complex. The calculated interaction energy indicates that the host molecule can hold two six membered aromatic guests at a time. The ΔG value also suggests that the inclusion of second guest is thermodynamically favorable. It is observed that in most cases the addition of second guest molecule decreases the interaction energy as well as ΔH and ΔG values compared to the Guest@ExBox$^{4+}$.4Cl$^{-}$ complex. It is obvious that this calculation provides a rough idea of interaction of the second guest molecule and a further analysis is needed to get more precise result.

## 3. Conclusion

The substitution of methine group (=CH–) of guest molecule with N atom decreases the extent of the host-guest interaction in solvent phase, but in gas phase no such regular change in interaction energy is observed. In solvent phase the interaction energy is also found to decrease with decreasing the NICS(0) value of the gas molecule. Even in presence of counter-ion (Cl$^{-}$) in most cases the interaction energy decreases with increasing the number of N atoms in the guest ring as well as decreasing the NICS(0) value of the guest molecule. The complexes are stabilized via π-type van der Waals interaction rather than charge transfer interaction. The interaction is visualized via corresponding 3D NCI plot. Since the azines and their derivatives are also used as antibiotic and antitumor agents hence the favorable interaction of ExBox$^{4+}$ with hetero cyclic aromatic compounds (azines) in addition to PAH may accelerate the possibility of using the new host complex in the field of medicinal chemistry.

**Acknowledgement:** PKC thanks the DST, New Delhi for the J. C. Bose National Fellowship. RD thanks UGC, New Delhi for a fellowship. We are thankful to Frieda Vansina for allowing us to use the trial version of ADF.

**Author Information**: Author for correspondence: Email: pkc@chem.iitkgp.ernet.in

Supporting Information: The 2D NCI plot and gas phase and solution phase geometry provided.

**Table 1.** NICS value for guest molecules, HOMO-LUMO gap (HLG, eV), interaction energy (ΔE, kcal/mol), zero point energy corrected interaction energy ($\Delta E_{ZPE}$ , kcal/mol), reaction enthalpy (ΔH, kcal/mol), reaction free energy (ΔG, kcal/mol) for Guest@$ExBox^{4+}$ complex in gas phase.

| No. of N atom | System | NICS (0) | NICS (1) | HLG (*eV*) | ΔE (kcal/mol) | $\Delta E_{ZPE}$ (kcal/mol) | ΔH (kcal/mol) | ΔG (kcal/mol) |
|---|---|---|---|---|---|---|---|---|
| 0 | **Benzene@$ExBox^{4+}$** | -8.963 | -11.324 | 7.573 | -19.9 | -18.6 | -18.5 | -7.4 |
| 1 | **Pyridine@$ExBox^{4+}$** | -7.531 | -11.096 | 7.492 | -18.9 | -17.8 | -17.3 | -5.9 |
| 2 | **Pyrimidine@$ExBox^{4+}$** | -6.053 | -10.687 | 7.490 | -18.9 | -17.4 | -17.2 | -4.2 |
| 2 | **Pyridazine@$ExBox^{4+}$** | -5.895 | -11.298 | 7.488 | -18.5 | -17.2 | -16.9 | -4.6 |
| 2 | **Pyrazine@$ExBox^{4+}$** | -5.829 | -11.098 | 7.613 | -17.0 | -16.0 | -15.5 | -3.9 |
| 3 | **1,2,3, Triazine@$ExBox^{4+}$** | -4.645 | -11.354 | 7.775 | -21.0 | -20.0 | -19.5 | -8.6 |
| 3 | **1,3,5Triazine@$ExBox^{4+}$** | -4.433 | -10.164 | 7.530 | -15.6 | -14.5 | -16.2 | -4.8 |
| 3 | **1,2,4 Triazine@$ExBox^{4+}$** | -3.956 | -10.959 | 7.551 | -19.5 | -18.3 | -18.0 | -5.1 |
| 4 | **1,2,3,4Tetrazine@$ExBox^{4+}$** | -2.826 | -11.419 | 7.473 | -21.6 | -20.8 | -24.3 | -13.1 |
| 4 | **1,2,4,5Tetrazine@$ExBox^{4+}$** | -1.753 | -11.040 | 7.608 | -18.2 | -17.1 | -16.7 | -5.4 |

**Table 2.** NICS value for guest molecules, interaction energy (ΔE, kcal/mol), reaction enthalpy (ΔH, kcal/mol), and reaction free energy (ΔG, kcal/mol) for Guest@$ExBox^{4+}$ complexes in solution phase

| No. of N atom | System | NICS(0) | | ΔE (kcal/mol) | | ΔH (kcal/mol) | | ΔG (kcal/mol) | |
|---|---|---|---|---|---|---|---|---|---|
| | | aceto | water | aceto | water | aceto | water | aceto | water |
| 0 | **Benzene@$ExBox^{4+}$** | -8.933 | -8.932 | -17.6 | -17.6 | -15.6 | -15.6 | -4.7 | -4.9 |
| 1 | **Pyridine@$ExBox^{4+}$** | -7.474 | -7.473 | -16.7 | -16.7 | -14.7 | -14.7 | -3.0 | -3.7 |
| 2 | **Pyrimidine@$ExBox^{4+}$** | -6.003 | -6.002 | -15.7 | -15.6 | -10.0 | -9.9 | 1.5 | 1.5 |
| 2 | **Pyridazine@$ExBox^{4+}$** | -5.798 | -5.798 | -15.9 | -16.1 | -14.1 | -14.2 | -1.6 | -1.3 |
| 2 | **Pyrazine@$ExBox^{4+}$** | -5.689 | -5.689 | -15.8 | -15.8 | -13.8 | -13.8 | -1.0 | -1.1 |
| 3 | **1,2,3, Triazine@$ExBox^{4+}$** | -4.389 | -4.388 | -15.3 | -15.2 | -13.7 | -13.7 | -1.7 | -1.7 |
| 3 | **1,3,5Triazine@$ExBox^{4+}$** | -4.310 | -4.309 | -14.2 | -14.2 | -12.6 | -12.6 | -1.8 | -1.9 |
| 3 | **1,2,4 Triazine@$ExBox^{4+}$** | -3.842 | -3.841 | -12.4 | -12.3 | -11.0 | -10.9 | 0.8 | 0.8 |
| 4 | **1,2,3,4Tetrazine@$ExBox^{4+}$** | -2.494 | -2.491 | -13.2 | -13.2 | -11.2 | -11.1 | 0.9 | 0.8 |
| 4 | **1,2,4,5Tetrazine@$ExBox^{4+}$** | -1.751 | -1.751 | -12.5 | -12.4 | -10.6 | -10.6 | 1.9 | 1.9 |

**Table 3.** Interaction energy (ΔE, kcal/mol), zero point energy corrected interaction energy ($\Delta E_{ZPE}$ , kcal/mol), reaction enthalpy (ΔH, kcal/mol), reaction free energy (ΔG, kcal/mol) for Guest@$ExBox^{4+}$ complex in presence of counter ion in gas phase

| No. of N atom | System | ΔE (kcal/mol) | ΔH (kcal/mol) | ΔG (kcal/mol) | $\Delta E_{ZPE}$ (kcal/mol) |
|---|---|---|---|---|---|
| 0 | **Benzene@$ExBox^{4+}.4Cl^-$** | -31.1 | -28.9 | -16.0 | -28.9 |
| 1 | **Pyridine@$ExBox^{4+}.4Cl^-$** | -29.6 | -28.0 | -15.5 | -28.0 |
| 2 | **Pyrimidine@$ExBox^{4+}.4Cl^-$** | -29.8 | -28.0 | -14.6 | -28.0 |
| 2 | **Pyridazine@$ExBox^{4+}.4Cl^-$** | -29.3 | -27.4 | -13.5 | -27.4 |
| 2 | **Pyrazine@$ExBox^{4+}.4Cl^-$** | -26.3 | -24.8 | -11.8 | -24.8 |
| 3 | **1,2,3, Triazine@$ExBox^{4+}.4Cl^-$** | -29.8 | -30.2 | -18.2 | -29.0 |
| 3 | **1,3,5Triazine@$ExBox^{4+}.4Cl^-$** | -29.6 | -28.1 | -16.0 | -28.1 |
| 3 | **1,2,4 Triazine@$ExBox^{4+}.4Cl^-$** | -29.3 | -27.7 | -14.3 | -27.7 |
| 4 | **1,2,3,4Tetrazine@ExBox$^{4+}$.4Cl$^{-}$** | -28.6 | -27.1 | -15.2 | -27.1 |
| 4 | **1,2,4,5Tetrazine@ExBox$^{4+}$.4Cl$^{-}$** | -17.3 | -16.6 | -5.7 | -16.6 |

**Table 4.** Energy decomposition analysis data for Guest@ExBox$^{4+}$ complex in gas phase

| System | $\Delta V_{eles}$ | $\Delta E_{pauli}$ | $\Delta E_{orb}$ | $\Delta E_{dis}$ | $\Delta E_{int}$ |
|---|---|---|---|---|---|
| **Benzene@ExBox$^{4+}$** | -23.8 (42.6%) | 28.0 | -11.6 (20.8%) | -20.4 (36.6%) | -27.8 |
| **Pyridine@ExBox$^{4+}$** | -20.4 (41.3%) | 23.6 | -10.9 (22.1%) | -18.2 (36.7%) | -25.9 |
| **Pyrimidine@ExBox$^{4+}$** | -28.9 (54.0%) | 24.6 | -10.0 (18.7%) | -14.6 (27.3%) | -28.9 |
| **Pyridazine@ExBox$^{4+}$** | -23.8 (41.7%) | 28.3 | -13.1 (22.9%) | -20.2 (35.4%) | -28.8 |
| **Pyrazine@ExBox$^{4+}$** | -16.8 (38.1%) | 19.1 | -9.7 (22.7%) | -16.3 (38.1%) | -23.7 |
| **1,2,3, Triazine@ExBox$^{4+}$** | -25.8 (50.6%) | 22.3 | -10.6 (20.9%) | -14.5 (28.5%) | -28.6 |
| **1,3,5Triazine@ExBox$^{4+}$** | -20.9 (48.5%) | 21.9 | -12.5 (29.0%) | -9.7 (22.5%) | -21.3 |
| **1,2,4 Triazine@ExBox$^{4+}$** | -28.8 (49.9%) | 30.5 | -11.8 (20.4%) | -17.1 (29.7%) | -27.1 |
| **1,2,3,4Tetrazine@ExBox$^{4+}$** | -29.7 (56.5%) | 21.9 | -11.0 (21.0%) | -11.9 (22.5%) | -30.7 |
| **1,2,4,5Tetrazine@ExBox$^{4+}$** | -29.7 (50.5%) | 32.5 | -11.9 (20.2%) | -17.3 (29.3%) | -26.5 |

**Table 5.** Charge decomposition analysis data for Guest@ExBox4+ complex in gas phase

| System | d | b | d-b | r |
|---|---|---|---|---|
| **Benzene@ExBox$^{4+}$** | 0.01477 | 0.02945 | -0.01469 | -0.14300 |
| **Pyridine@ExBox$^{4+}$** | 0.01589 | 0.03106 | -0.01517 | -0.11723 |
| **Pyrimidine@ExBox$^{4+}$** | 0.00772 | 0.04567 | -0.03794 | -0.12722 |
| **Pyridazine@ExBox$^{4+}$** | 0.01508 | 0.04065 | -0.02557 | -0.13786 |
| **Pyrazine@ExBox$^{4+}$** | 0.01760 | 0.03574 | -0.01814 | -0.09666 |
| **1,2,3, Triazine@ExBox$^{4+}$** | 0.00557 | 0.04006 | -0.03448 | -0.09134 |
| **1,3,5Triazine@ExBox$^{4+}$** | 0.01411 | 0.03566 | -0.02155 | -0.12043 |
| **1,2,4 Triazine@ExBox$^{4+}$** | 0.00505 | 0.04546 | -0.04041 | -0.13450 |
| **1,2,3,4Tetrazine@ExBox$^{4+}$** | 0.00201 | 0.03874 | -0.03672 | -0.08288 |
| **1,2,4,5Tetrazine@ExBox$^{4+}$** | 0.00523 | 0.04029 | -0.03506 | -0.14592 |

**Table 6.** HOMO-LUMO gap (HLG, eV), interaction energy (ΔE, kcal/mol), reaction enthalpy (ΔH, kcal/mol), reaction free energy (ΔG, kcal/mol) for Guest@ExBox4+ complex in presence of counter ion in gas phase

| System | HLG (eV) | Sequential | | | Average | | |
|---|---|---|---|---|---|---|---|
| | | ΔE (kcal/mol) | ΔH (kcal/mol) | ΔG (kcal/mol) | ΔE (kcal/mol) | ΔH (kcal/mol) | ΔG (kcal/mol) |
| **2Benzene@ExBox$^{4+}$** | 5.382 | -27.4 | -25.4 | -13.2 | -23.8 | -22.1 | -10.4 |
| **2Pyridine@ExBox$^{4+}$** | 5.163 | -26.6 | -24.5 | -12.0 | -23.7 | -21.5 | -8.5 |
| **2Pyrimidine@ExBox$^{4+}$** | 5.182 | -27.6 | -25.6 | -12.5 | -25.8 | -24.2 | -11.5 |
| **2Pyridazine@ExBox$^{4+}$** | 5.216 | -33.0 | -30.5 | -15.8 | -39.8 | -36.5 | -19.8 |
| **2Pyrazine@ExBox$^{4+}$** | 5.194 | -26.5 | -24.7 | -11.9 | -23.3 | -21.8 | -9.1 |
| **2(1,2,3, Triazine)@ExBox$^{4+}$** | 5.424 | -32.2 | -30.0 | -17.8 | -32.7 | -30.3 | -17.3 |
| **2(1,3,5Triazine)@ExBox$^{4+}$** | 5.393 | -22.7 | -21.1 | -9.1 | -16.1 | -14.7 | -2.1 |
| **2(1,2,4 Triazine)@ExBox$^{4+}$** | 5.007 | -26.2 | -24.2 | -11.2 | -22.8 | -21.1 | -8.2 |
| **2(1,2,3,4Tetrazine)@ExBox$^{4+}$** | 5.539 | -26.6 | -24.5 | -11.3 | -24.6 | -22.5 | -7.3 |
| **2(1,2,4,5Tetrazine)@ExBox$^{4+}$** | 5.318 | -19.6 | -18.3 | -6.9 | -21.9 | -20.5 | -8.1 |

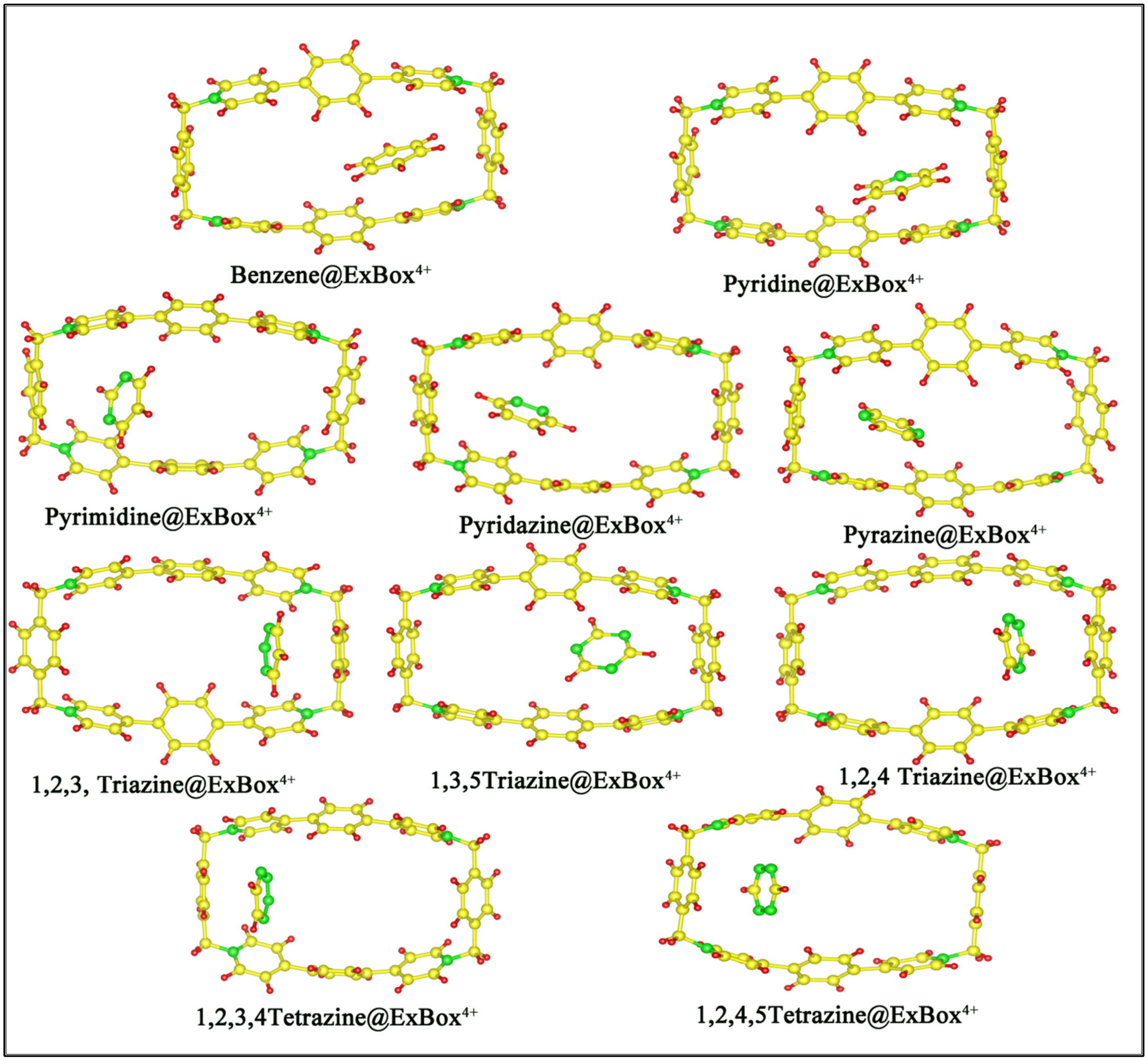

**Figure 1**. The geometry of Guest@ExBox$^{4+}$ complexes

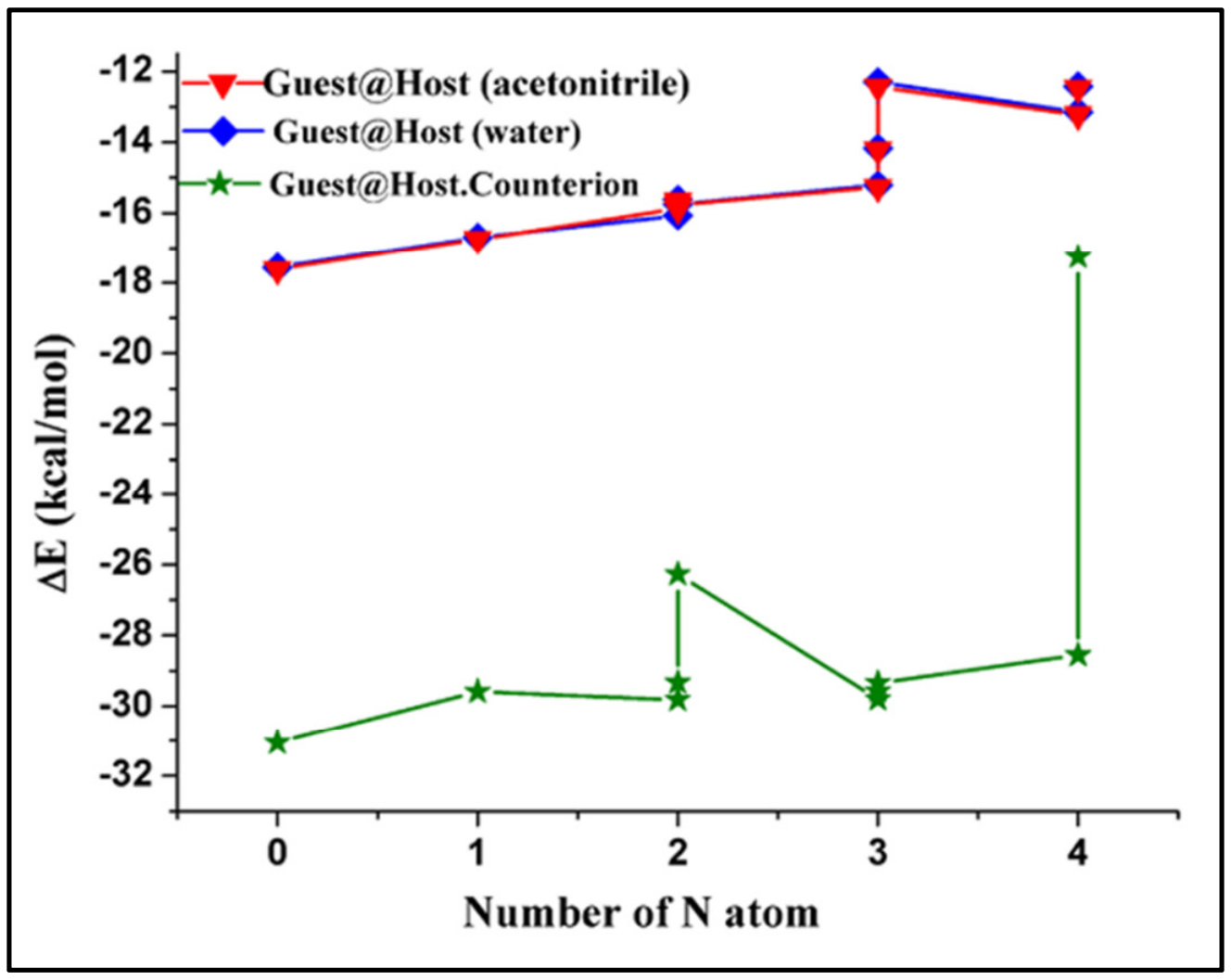

**Figure 2.** Variation of interaction energy with number of N atom present in guest ring

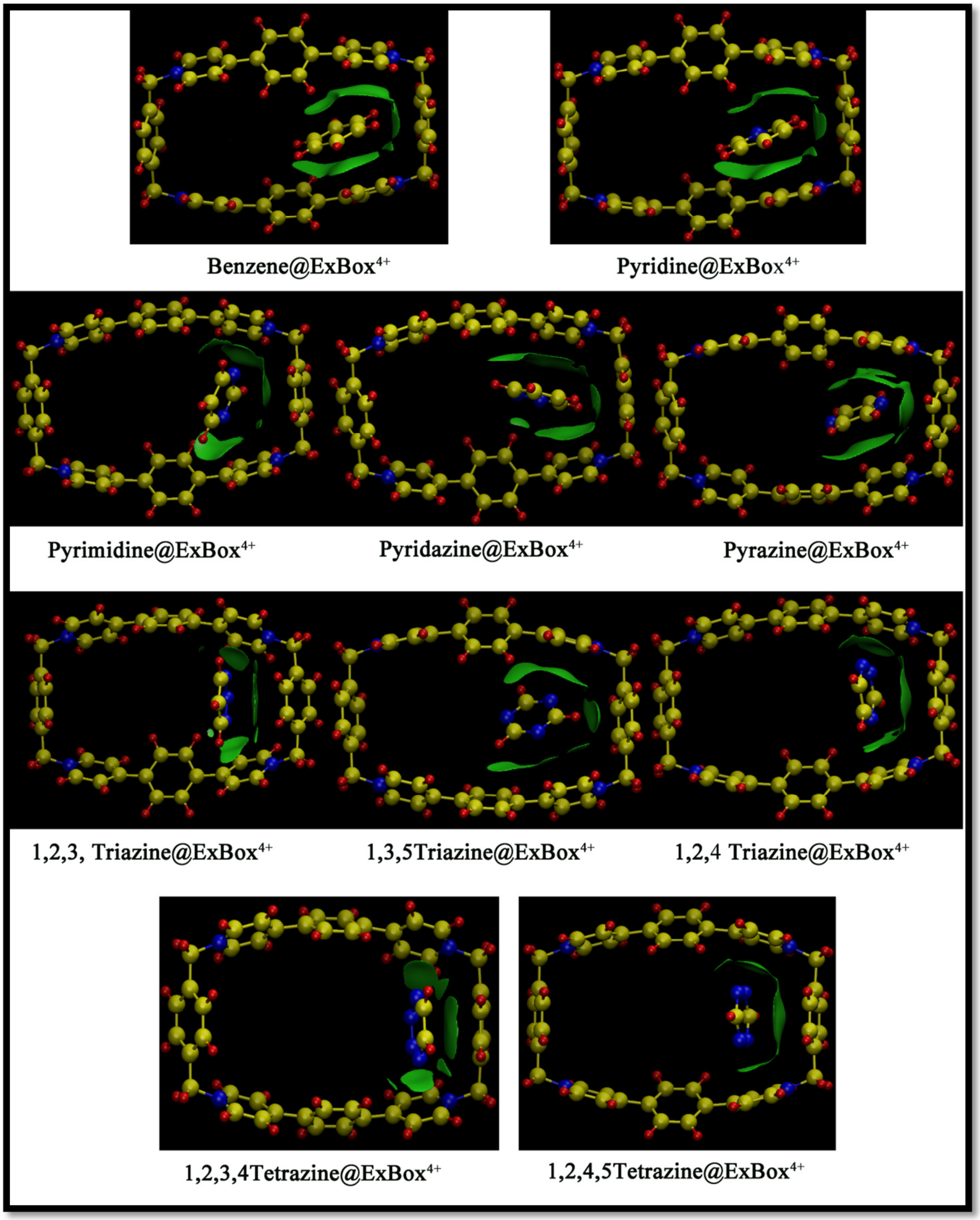

**Figure 3**. 3D plots of NCI between guest and $ExBox^{4+}$ complex. NCI surface shows the intermolecular interaction, gradient cutoff is $s = 0.4$ au and the color scale is $-0.05 < \rho < 0.05$ au